\documentclass{PoS}

\usepackage{amsmath}

\title{Discovery of VHE gamma-rays from the vicinity of the shell-type SNR
G318.2+0.1 with H.E.S.S.}

\ShortTitle{Discovery of VHE gamma-rays from the vicinity of the shell-type SNR
G318.2+0.1 with H.E.S.S.}

\author{P. Hofverberg$^a$, R.C.G. Chaves$^a$, A. Fiasson$^b$, K. Kosack$^c$, J. M\'ehault$^d$ and E. de On\~a Wilhelmi$^a$ for the H.E.S.S. Collaboration\\
\llap{$^a$} MPIK Heidelberg, Germany\\
\llap{$^b$} CNRS/IN2P3, Universit\'e de Savoie, Annecy-le-Vieux, France\\
\llap{$^c$} IRFU/DSM/CEA, CE Saclay, France\\
\llap{$^d$} CNRS/IN2P3, Universit\'e de Montpellier 2, Montpellier, France\\
E-mail: \email{petter.hofverberg@mpi-hd.mpg.de}}
\abstract{The on-going H.E.S.S. Galactic Plane Survey continues to reveal new sources of VHE $\gamma$-rays. In particular, recent re-observations of the region around the shell-type supernova remnant (SNR) G318.2+0.1 have resulted in the discovery of statistically-significant very-high-energy (VHE) $\gamma$-ray emission from an extended region. Although the source remains unidentified, archival observations of $^{12}$CO in the region provide an opportunity to investigate a potential SNR/molecular cloud interaction. The morphological properties of this newly-discovered VHE $\gamma$-ray source HESS$\,$J1457$-$593 are presented and discussed in light of the multi-wavelength data available.}

\FullConference{25th Texas Symposium on Relativistic Astrophysics \\
                 December 6-10, 2010\\
                 Heidelberg, Germany}

\begin{document}

\section{Introduction}
\label{sec:introduction}
Supernova remnants (SNR) are thought to be responsible for the acceleration of cosmic rays (CRs) up to energies around the ``knee'' ($\sim$$10^{15}$~eV) of the CR spectrum, a statement which is backed up by both theoretical arguments and experimental evidence. Theoretically, it is well established that supernova explosions release just the right amount of energy into the interstellar medium to account for the energy budget of the CRs, assuming they can convert on the order of $\sim$10\% of their energy into kinetic energy of CRs \cite{ginzburg}. Furthermore, models exist that can explain how CRs can be accelerated through diffusive shock acceleration at the shock front up to energies approaching the knee. Experimental evidence comes from primarily X-ray and very-high-energy (VHE; E > 0.1 TeV) $\gamma$-ray studies of young shell-type SNRs such as SN 1006 \cite{Koyama1995255}\cite{hessSN1006}, which have found evidence for non-thermal populations of CRs extending up to $\sim$100~TeV. 

$\gamma$-rays can be produced in both hadronic interactions (with subsequent pion decay) and from energetic electrons that inverse Compton scatter off background photon fields. An ambiguity therefore exists in the responsible radiating particle population when detecting $\gamma$-rays from SNRs, and determining the hadronic or leptonic nature of the accelerating mechanism remains a key issue. Since neutrino detectors have not yet proven to be sensitive enough to detect neutrinos from astrophysical sources, no direct indicator of either process currently exists. However, disentangling the radiation processes can be facilitated by studying SNR $\gamma$-ray spectra extending well beyond 10~TeV, where inverse-Compton spectra tend to cut off due to strong radiative energy losses of the parent electron populations, or by studying SNRs interacting with dense molecular clouds where the target material in the cloud could boost the $\gamma$-ray component from a hadronic mechanism.

For the reasons outlined above, the SNR G318.2+0.1 may be a prime target for studying VHE $\gamma$-ray emission and the radiation mechanism that generates the emission. G318.2+0.1 is a large diameter (40$\prime\prime\times$35$\prime\prime$) shell-type SNR discovered in the Molonglo Observatory Synthesis Telescope (MOST) survey \cite{1996A&AS..118..329W} and is characterized by two non-thermal filaments in the Northwest and Southeast which form two sections of a shell, and a central region of thermal emission which corresponds to a HII region. 

In this paper, the discovery of the VHE source HESS$\,$J1457$-$593 is presented and discussed in light of multiwavelength data that suggest that the emission originates from an interaction between SNR G318.2+0.1 and a molecular cloud.

\section{H.E.S.S. observations and Analysis}

\subsection{The H.E.S.S. Telescope Array}
The High Energy Stereoscopic System (H.E.S.S.) is an array of Cherenkov telescopes for VHE $\gamma$-ray astronomy by observing Cherenkov light emitted from $\gamma$-ray induced extensive air showers (EASs). H.E.S.S. is comprised of four identical 12~m diameter telescopes arranged in a square with 120~m sides and is located in the Khomas highlands of Namibia in the southern hemisphere ($23^{\circ}16'17'' \text{S}$) at a height of 1800~m above sea level. The array has been fully operational since 2004~\cite{hesscrabpaper}. Each telescope is equipped with an alt-az mount and a reflector of Davis-Cotton design with a total mirror area of 107~m$^2$ and a focal length of 15~m~\cite{hessopticalpaper}. A photomultiplier camera with 960 pixels (each of $0.16^{\circ}$ size) provides a total field-of-view of $5^{\circ}$ which makes H.E.S.S. well suited for observing extended sources and for performing survey observations. The H.E.S.S. array uses a stereoscopic trigger which drastically reduces the background from single muons and the night sky, keeping dead-time low and allowing a low energy threshold. In addition, employing the stereoscopic detection technique, an angular resolution of $\sim$$0.1^{\circ}$ and an energy resolution of $\sim$15\% is achieved. H.E.S.S.'s unprecedented sensitivity to $\gamma$-rays permits it to detect a point source with a flux of 1\% of the Crab Nebula at a significance of 5~$\sigma$ in only $\sim$25~h of observations~\cite{hesscrabpaper}.

% maybe merge the next two subsections into: "Data and Analysis Methods"
\subsection{Data and Analysis Methods}
The region of interest was observed with the H.E.S.S. telescope array between 2004 and 2010. The data-set is primarily comprised not of dedicated observations of the VHE $\gamma$-ray source presented in this paper but rather of survey observations of the region and from observations of nearby sources. After standard quality selection \cite{hesscrabpaper} to remove observations taken during unfavorable weather conditions or affected by hardware-related problems, the live-time amounts to $\sim$102~h. The effective live-time at the source in question is however significantly lower due to the large average offset of the pointings relative to the source ($1.6^{\circ}$). The data was taken in a series of runs of typical duration $\sim$28 min at zenith angles between $35^{\circ}$ to $55^{\circ}$, with a mean of $38^{\circ}$.

The data-set was analyzed using the Hillas second moment method \cite{hillas1985} employed in the H.E.S.S. standard analysis \cite{hesscrabpaper}. To separate $\gamma$-ray event candidates from CR-like events, \emph{hard cuts} were used, where a minimum of 200 photoelectrons is required in each recorded EAS image. Compared to \emph{standard cuts}, which requires a minimum of 80 photoelectrons, this results in a narrower PSF and an improved background rejection, but also in an increased energy threshold. The time-dependent optical response of the telescopes, due to mirror deterioration, was calibrated using the known amount of Cherenkov light from single muons passing close to the telescope~\cite{hesscrabpaper}.

For the generation of 2D images, the \emph{ring background method} \cite{2007A&A...466.1219B} was used, where the background at each point in the sky was calculated within an annulus centered on the point in question, with an inner radius of $0.7^{\circ}$ and a area factor between the OFF and ON region of 7. The statistical significance of the images were derived from the number of off-source (background) and on-source events, following the likelihood ratio procedure in \cite{1983ApJ...272..317L}. All results presented in this paper were cross-checked with the Model2D analysis \cite{2006astro.ph..7247D} which is based on a model of Cherenkov image parameters and also utilizes an independent calibration of the raw data. Consistent results were obtained with the two analysis techniques.

%_________________________________________________________________

\section{Results}
\begin{figure}
  \centering
  \includegraphics[width=0.75\textwidth]{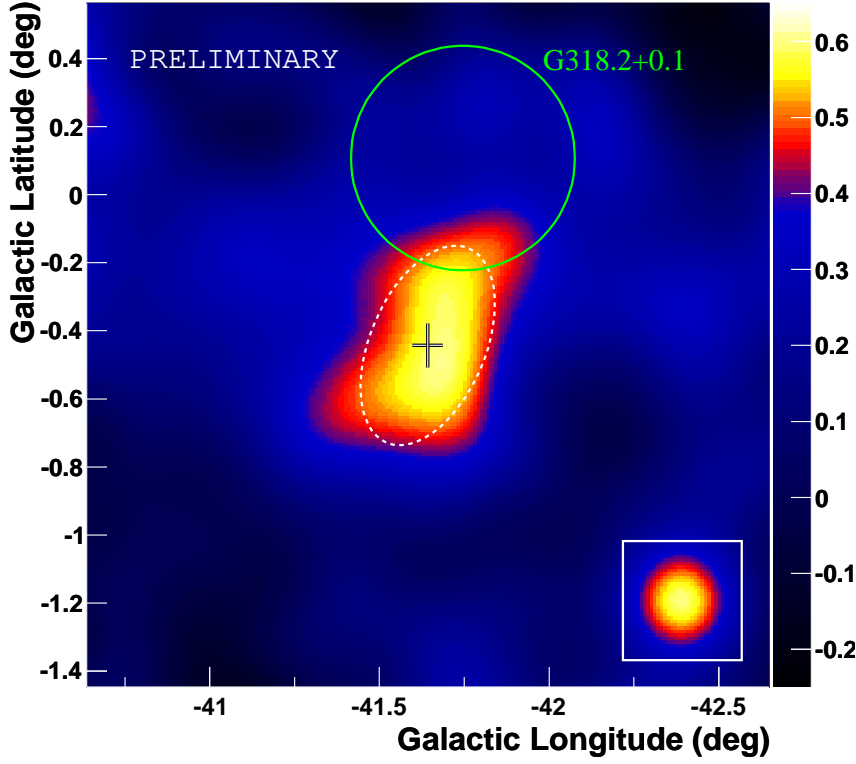}
  \caption{Image of the VHE $\gamma$-ray excess (in units of $\gamma$-rays arcmin$^{-2}$) of HESS$\,$J1457$-$593, corrected for exposure and smoothed with a 2D Gaussian with a width of $\sigma=0.12^{\circ}$. The color scale is set so that the blue/red color transition occurs at the level of statistically significant features ($5\sigma$). The black and white cross indicates the best-fit position of the source centroid, and its size represent the statistical errors of the fit. The fitted extension of the source is indicated by the white dashed ellipse. The inset in the lower right corner shows the point spread function for this particular data-set, smoothed in an identical fashion as the main figure. The green circle shows the approximate extent of the SNR G318.2+0.1 in radio.}
\label{fig:hessexcess}
\end{figure}

Figure~\ref{fig:hessexcess} shows the gamma-ray excess counts in a $2^{\circ}\times2^{\circ}$ region around the source of interest. The image has been smoothed with a Gaussian kernel with standard deviation $0.12^{\circ}$. The detection significance of the source amounts to 8.9~$\sigma$ using an integration radius of $0.4^{\circ}$. Table~1 gives a summary of the event statistics from the analysis. The source centroid and extension were determined by fitting an asymmetric 2D Gaussian, convolved with the H.E.S.S. point-spread function, to the unsmoothed $\gamma$-ray excess map. The best-fit centroid is at $\alpha_{\text{J}2000}=14^{\text{h}}57^{\text{m}}46^{\text{s}}\pm33^{\text{s}}$ and $\delta_{J2000}=-59^{\circ}28^{\prime}\pm 5^{\prime}$ ($\text{l}\sim 318.3^{\circ}$, $\text{b}\sim -0.4^{\circ}$), giving a $\chi^2/\text{ndf}=778.7/778$, and the source is thus named HESS$\,$J1457$-$593. The best-fit position is indicated in Figure~\ref{fig:hessexcess} by a white cross, the extension of which corresponds to the statistical error of the fit (quoted above). The best-fit source extension is $\sigma_{\text{major}}=0.31^{\circ}\pm0.07^{\circ}$, $\sigma_{\text{minor}}=0.17^{\circ}\pm0.05$, at an angle $\phi=67^{\circ}\pm 15^{\circ}$. It should be noted that due to the non-Gaussian morphology of the source, this procedure only gives a rough estimate of the source centroid and extension. The north part of HESS$\,$J1457$-$593 overlaps the southern part of the SNR G318.2+0.1, whose approximate extension in the radio is indicated in Figure~1 by a green circle.

\begin{table}[h]
\centering
\begin{tabular}{cccccccc}
Source Name & $\alpha_{\text{J}2000}$ & $\delta_{\text{J}2000}$ & $\text{N}_{\text{on}}$ & $\text{N}_{\text{off}}$ & $\alpha$ & Excess & Significance $[\sigma]$ \\
\hline
HESS$\,$J1457$-$593 & $14^{h}57^{m}46^{s}$ & $-59^{\circ}28^{\prime}$ & 5034 & 22547 & 0.19 & 659 & 8.9\\
\label{tab:eventstatistics}
\end{tabular}
\caption{HESS$\,$J1457$-$593 source centroid (columns 2 and 3), event statistics (columns 4-7) and peak significance (column 8).}
\end{table}

\section{Observations of molecular clouds}
\begin{figure}
  \centering
  \includegraphics[width=0.85\textwidth]{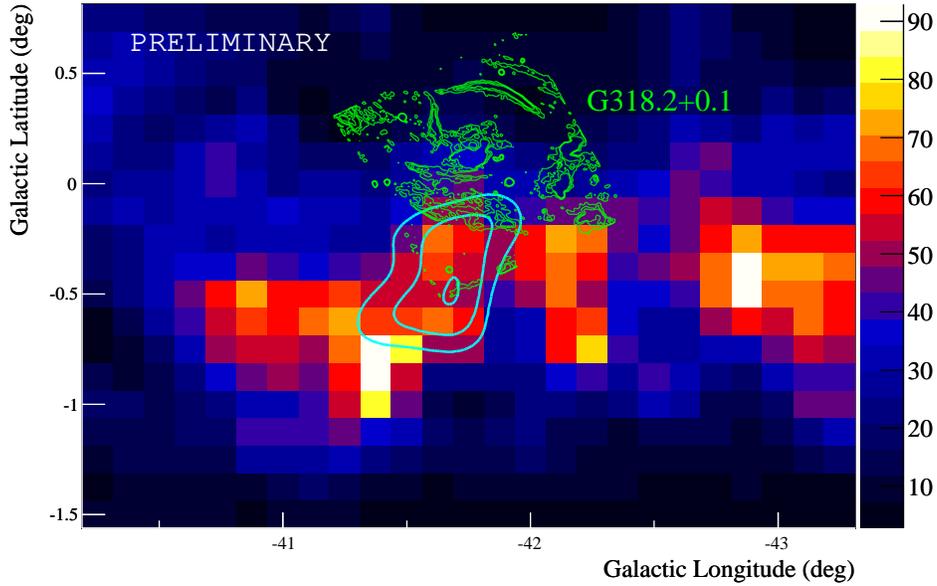}
  \caption{Map of the $^{12}$CO (J:1$\rightarrow$0) line emission around HESS$\,$J1457$-$593 integrated between $-54.5$~km/s and $-32.5$~km/s. Overlaid in turquoise are the excess count contours of HESS$\,$J1457$-$593 (corresponding to 0.40, 0.50 and 0.60 $\gamma$-rays arcmin$^{-2}$). In green are the radio contours (843~MHz) of the SNR G318.2+0.1. The contours correspond to intensities of 2, 6 and 10 mJy$/$beam. The colorbar units are in K$\,$km/s.}
  \label{fig:co12map}
\end{figure}

In searching for molecular cloud counterparts, $^{12}$CO data from the Dame survey \cite{2001ApJ...547..792D} was analyzed. A square box, $1^{\circ}\times 1^{\circ}$ large, centered on the best fit position of HESS$\,$J1457$-$593 was used as an integration region to search for evidence of molecular clouds in the region of the H.E.S.S. source. Integrating $^{12}$CO over the full Local Standard of Rest velocity ($V_{\text{LSR}}$) range, a strong peak was found in the velocity interval $-54.5$~km/s to $-32.5$~km/s centered at V$_{LSR}=-42.0$~km/s. A map of the $^{12}$CO line emission integrated in this velocity interval is shown in Figure~\ref{fig:co12map}. The color transition is chosen so that the blue-red transition start at the boundaries of the molecular cloud, derived using the definition of molecular cloud boundaries suggested by \cite{seta1998}. The cloud complex coincident with the H.E.S.S. source measures roughly $1.8^{\circ}\times1.1^{\circ}$ and fully contains the VHE $\gamma$-ray emission (indicated by black contours).

The distance to the molecular cloud is estimated assuming the Galactic rotation curve from \cite{nakanishi2003} with a rotation speed of V$_0=217$~km/s  and a Galactocentric distance of the sun of R$_0=8.5$~kpc. A velocity distance of $-42.0$~km/s then corresponds to a near/far kinematic distance of ($3.5\pm0.2$)~kpc and ($9.2\pm0.2$)~kpc respectively, where the errors are estimated by comparing the results above with the results obtained using a Galactic rotation model from \cite{burton1978}\footnote{The distance uncertainty arising from the velocity resolution of the $^{12}$CO survey, 1.3~km/s, is negligible.}. 

Assuming a relationship between the velocity-integrated CO intensity W$_{\text{CO}}$ and the molecular hydrogen column density N($\text{H}_2$) as  N$(\text{H}_2)$/W$_{\text{CO}}=1.8\times10^{20}$ \cite{2001ApJ...547..792D}, the cloud properties given in Table~2 are derived for the near kinematic distance solution. Using the far distance solution, the molecular cloud mass would be in the order of $10^{7}$~M$_{\odot}$. Such massive clouds have not been observed outside the Galactic centre and the near distance solution is therefore adopted throughout this paper. In the above calculations, it is assumed that the cloud extension in the line-of-sight is equal to the average cloud extension perpendicular to the line-of-sight. The molecular cloud coincident with the H.E.S.S. source has the characteristics of a typical Giant Molecular Cloud (GMC) with a mass on the order of $10^{5}$~M$_{\odot}$ and an extension of tens of parsec \cite{solomon1979}. 

\begin{table}[h]
\centering
\begin{tabular}{ccccc}
Position             & Angular Extension             & Average Extension & Mass                       & Density \\
\hline
(l,b)=$\sim$(318.4$^{\circ}$,-0.5$^{\circ}$)   & $1.8^{\circ}\times1.1^{\circ}$ & $\sim$80~pc             & $\sim$$3\times 10^{5}\,\text{M}_{\odot}$ &  $\sim$40~cm$^{-3}$
\label{tab:cloudprop}
\end{tabular}
\caption{Derived quantities of the GMC coincident with the source HESS$\,$J1457$-$593. The first column gives the geometrical centre position of the cloud, the second its angular extension (in Galactic longitude, latitude). Columns three to five gives the average (physical) extension, the mass and the density of the cloud, all assuming a distance to the cloud of 3.5~kpc.}
\end{table}

\section{Discussion and Conclusions}
An extended source of VHE $\gamma$-ray emission, HESS J1457-593, has been found in the vicinity of the SNR G318.2+0.1. As shown in Figure~\ref{fig:co12map}, the TeV $\gamma$-ray emission is partially overlapping the non-thermal, southern rim of the SNR and has an elongated morphology. A GMC complex is positionally coincident with the TeV $\gamma$-ray emission and overlapping the southern rim of the SNR, suggesting a region of cosmic-ray enhancement and providing an excellent opportunity to investigate a putative SNR-GMC interaction and to study cosmic ray propagation through a molecular cloud.

Assuming the GMC and the SNR G318.2+0.1 are spatially associated, the distance to the SNR, unknown until now, is given. This is a key parameter of SNRs and allows an estimation of its physical size and age. At a distance of 3.5~kpc and with an angular size of (40$\prime\prime\times$35$\prime\prime$), the SNR would have a physical diameter of roughly 40~pc. A Sedov-Taylor model of SNR evolution \cite{blondin1998} is then used to estimate its age directly from the SNR diameter and the ambient density which can be assumed to be 1~cm$^{-3}$, the average for Galactic SNRs. This gives an age of SNR G318.2+0.1 of roughly 8000~years\footnote{The age does not depend strongly on the ambient density. A rough approximation for this particular case is: $\text{t}=(5+3\times\text{n})\text{kyr}$.}. SNR G318.2+0.1 would then be in the late Sedov phase and could thus be an efficient particle accelerator.

\acknowledgments
The support of the Namibian authorities and of the University of Namibia in facilitating the construction and operation of H.E.S.S. is gratefully acknowledged, as is the support by the German Ministry for Education and Research (BMBF), the Max Planck Society, the French Ministry for Research, the CNRS-IN2P3 and the Astroparticle Interdisciplinary Programme of the CNRS, the U.K. Science and Technology Facilities Council (STFC), the IPNP of the Charles University, the Polish Ministry of Science and Higher Education, the South African Department of Science and Technology and National Research Foundation, and by the University of Namibia. We appreciate the excellent work of the technical support staff in Berlin, Durham, Hamburg, Heidelberg, Palaiseau, Paris, Saclay, and in Namibia in the construction and operation of the equipment.

\end{document}